\documentclass[12pt]{article}

\usepackage[margin=1in]{geometry}
\usepackage{amsmath}
\usepackage{amssymb}
\usepackage{graphicx}
\usepackage{booktabs}
\usepackage{array}
\usepackage{longtable}
\usepackage{natbib}
\usepackage{setspace}
\usepackage{caption}
\usepackage{subcaption}
\usepackage{float}
\usepackage{xcolor}
\usepackage{url}
\usepackage[ruled,vlined]{algorithm2e} 
\usepackage{comment}
\usepackage{tabularx}
\newcolumntype{Y}{>{\raggedright\arraybackslash}X}
\usepackage{adjustbox}
\usepackage{multirow}
\usepackage{newtxtext,newtxmath}
\usepackage{hyperref}

\graphicspath{{figures/}}

\hypersetup{
    colorlinks=true,
    linkcolor=blue,
    citecolor=blue,
    urlcolor=blue
}

\begin{document}


\begin{center}

{\LARGE\bfseries
\linespread{0.95}\selectfont
Design and Validation of a Grid-based Home\\
Detection via Stay-Time (GHOST) Software for\\
Mobile Location Data
\par}

\vspace{1.0cm}

{\large
Alessandra Recalde\textsuperscript{a,*},
Mustafa Sameen\textsuperscript{a},
Xiaojian Zhang\textsuperscript{a,b},
and Xilei Zhao\textsuperscript{a}
\par}

\vspace{0.5cm}

{\large
\textsuperscript{a}Department of Civil and Coastal Engineering,
University of Florida, USA
\par}

\vspace{0.12cm}

{\large
\textsuperscript{b}Kittelson \& Associates, Inc.,
Orlando, Florida, 32801 USA
\par}

\end{center}

\vspace{1.1cm}

\noindent
\textbf{*Corresponding author:}
Department of Civil and Coastal Engineering, University of Florida,\\
1949 Stadium Road, Gainesville, Florida 32611, USA.\\
\textbf{E-mail address:}
\href{mailto:arecalde@ufl.edu}{arecalde@ufl.edu}
(A. Recalde).

\vspace{0.35cm}


\vspace{0.65cm}

\section{Abstract}

Accurately detecting home locations from GPS data generated by mobile devices is a foundational step in human mobility research, with significant implications for transportation planning, public health, and emergency response. However, existing home detection algorithms often produce unreliable results for noisy real-world data and are barely validated due to a lack of ground-truth benchmarks. To tackle these limitations, this study presents the development and validation of a Grid-based home detection via Stay-Time (GHOST) algorithm, implemented as an open-source Python package. The algorithm infers proxy home locations by identifying the most frequently visited nighttime or weekend daytime grid cells based on customizable spatial and temporal filters. To validate its performance, we use the large-scale BostonWalks dataset, which includes over 155,000 trips from 377 participants in the Boston metropolitan area, to test robustness to noisy data. Additionally, we collected a ground-truth dataset for ten volunteers across different regions in the U.S., including Florida, Mississippi, and Colorado, along with their self-reported home coordinates, to evaluate GHOST across diverse mobility patterns and sampling conditions. We compared GHOST accuracy to that of 5 well-established home detection algorithms: All-time clustering method, Stay-point method, DBSCAN, K-MEANS++, and SciKit-Mobility Home Detection, across multiple parameter settings. Results show that GHOST outperforms all algorithms in accuracy and robustness, with average errors as low as 22.3 meters under optimal configurations. Our findings highlight the high accuracy and flexibility of our algorithm, with grid size being the most influential parameter during validation, demonstrating the potential of this algorithm for real-world mobile location data analysis.



\section{Introduction}

Large-scale mobile device location data has become a critical tool for understanding human mobility across urban, regional, and global contexts. These datasets, often collected via GPS-enabled mobile apps and distributed through commercial vendors such as SafeGraph, Veraset, and Cuebiq, have enabled researchers to analyze human movement at unprecedented temporal and spatial resolution \citep{hu2021, cheng2025mobilephone, barbosa2018}. From transportation planning and public health to disaster response and urban analytics, mobility data has fueled wide-ranging advancements across disciplines \citep{huang2022, barbosa2018,  hu2021, bian2024, sadeghinasr2019a2, jiang2025}. 

A foundational task in mobility research is home location detection—the process of inferring an individual’s proxy residence from patterns in their GPS traces. Accurate home detection is essential because it serves as the starting point for most mobility analyses. It anchors origin-destination flows, determines how individuals are linked to socio-demographic characteristics through spatial joins with census or administrative boundaries, and informs classifications of daily behavior such as commuting patterns or activity space \citep{jiang2025, sadeghinasr2019a2, chen2017}. For example, identifying home locations allows us to model trips and quantify human mobility metrics, like trip duration, travel mode, and travel purpose \citep{demissie2018trip, song2010modelling}. Furthermore, in evacuation and migration studies, knowing an individual's home locations before an emergency is necessary to determine evacuation routes, status, and overall mobility patterns \citep{yabe2019cross, Yabe2022ATC,lai2019exploring}. Determining home locations is foundational to a wide array of applications. Inaccuracies at this stage can lead to biased or misleading results, distorting travel models, misallocating resources, and weakening the policy relevance of findings \citep{wang2025}. 

However, most home detection algorithms rely on simplifying behavioral assumptions, including nighttime activity patterns and stable spatial clustering, which may introduce bias for individuals with irregular schedules or unevenly sampled data \citep{verma2024, gallotti2024, vanhoof2020}. These methods typically define home as the most frequent nighttime location (e.g., 6 PM to 7 AM), which can exclude valid observations and introduce bias for individuals with nontraditional schedules or irregular device usage \citep{huang2022, dabiri2018}. Additionally, many approaches rely on clustering and density-based methods, which group points based on spatial proximity and density while often neglecting temporal dynamics, making it difficult to distinguish transient visits from meaningful activity locations \citep{chen2014, wang2023}. These methods are sensitive to parameter selection, struggle with heterogeneous densities, and are vulnerable to noise and sparse sampling, where high-density regions may not correspond to true activity centers and low-frequency data can degrade performance \citep{luo2017, ester1996, zheng2010}. These limitations demonstrate the challenges of applying traditional clustering-based approaches to noisy, irregular, and temporally complex GPS mobility data, while the lack of ground-truth benchmarks further constrains evaluation and comparison across methods.

To address these limitations, this paper presents the development and validation of a \textbf{Grid-based home detection via Stay-Time (GHOST)} algorithm. Rather than relying solely on visit frequency or simple point counts, which can be misleading due to GPS sampling variations, brief visits, or noise bursts, GHOST introduces a hierarchical selection criterion that prioritizes \textit{stay-time}, the total duration spent in each grid cell, as the primary metric for home inference. The algorithm projects GPS points into a metric coordinate system, filters them by configurable nighttime hours, and overlays a spatial grid across the study area. GHOST identifies the home location as the grid cell with the greatest stay-time, using total GPS point count as a tie-breaker when necessary, and then further divides the winning cell into smaller bins to identify the densest area of activity. This approach relies on efficient linear-time spatial aggregation and includes fallback mechanisms for sparse data, making it robust to GPS noise, outliers, and irregular observations.

What distinguishes GHOST is its \textbf{adaptive temporal filtering}, which includes a built-in weekend fallback mechanism when nighttime data are unavailable. This flexibility allows the algorithm to infer homes using either nighttime or weekend daytime observations, improving applicability across users with atypical mobility patterns or inconsistent device usage.

To validate GHOST, we evaluate its performance using both the BostonWalks dataset from MIT and a controlled GeoTracker dataset, enabling assessment under both high-quality and real-world sparse mobility data conditions. Using this evaluation framework, we systematically compare GHOST against five widely used home detection algorithms—All-time clustering (A1), Stay-point (A2), DBSCAN, K-Means++, and SciKit-mobility home detection (SciKit)—to quantify accuracy and identify common methodological failure modes.

The remainder of this paper is organized as follows: Section 2 reviews relevant literature on home detection algorithms, introduces the 6 algorithms that will be explored in this study, and discusses their strengths and limitations. Section 3 describes the design and computational logic of the GHOST algorithm in detail. Section 4 introduces the ground-truth and BostonWalks datasets and outlines the data collection methodology. Section 5 introduces the sensitivity analysis and details the parameter variations tested across each algorithm. Section 6 reports findings from our sensitivity tests across algorithm parameters. Section 7 discusses key findings and implications, and Section 8 concludes with a summary and future directions.


\section{Literature Review}

Home location detection is a fundamental task in human mobility research as it provides the basis for applications such as origin-destination modeling, activity classification, exposure assessment, and demographic inference \citep{verma2024}. It supports a wide range of use cases across transportation planning, public health, urban analytics, and disaster response \citep{barbosa2018, huang2022}, and provides a critical reference point for interpreting travel behavior and commuting patterns \citep{schneider2013}. With the increasing availability of large-scale GPS data from mobile devices, the field has shifted from survey-based approaches toward automated, data-driven methods that infer home locations directly from spatiotemporal trajectories \citep{zandbergen2009}. Consequently, many home detection algorithms have emerged, each differing in how spatial and temporal information are utilized.

A common foundation across these methods is the assumption that individuals spend a disproportionate amount of time at or near their residence, particularly during nighttime hours \citep{ahas2010}. Many algorithms operationalize this assumption by identifying the most frequently visited location during evening and overnight periods (e.g., 6 PM to 7 AM) over extended observation windows \citep{huang2022}. While effective for individuals with regular daily schedules, this assumption can break down for populations with nontraditional routines, such as night-shift workers or individuals with multiple residences, as well as for users with sparse or irregular data, introducing uncertainty in inferred home locations \citep{dabiri2018, verma2024}.

Many approaches also rely on clustering and density-based techniques that identify meaningful locations through spatially dense groupings of points. These methods implicitly assume that activity locations form well-defined, compact clusters and that higher point density corresponds to behavioral significance. However, such assumptions are often violated in GPS data, where human mobility patterns are irregular, spatial distributions are uneven, and repeated movement through a location can artificially inflate density without indicating true activity \citep{ester1996, zheng2010}. Clustering algorithms are also sensitive to parameter selection and struggle with heterogeneous densities, where a single parameter set cannot accurately capture both dense and sparse regions \citep{schubert2017, luo2017}. Additionally, noise and irregular sampling can distort cluster boundaries or fragment meaningful locations, further reducing reliability. As a result, dense clusters may reflect sampling artifacts rather than true activity centers, limiting the effectiveness of clustering-based home detection approaches.

In contrast, grid-based methods offer a more flexible and robust alternative by discretizing space into fixed spatial units, thereby avoiding assumptions about cluster shape and reducing sensitivity to noise and sampling variability \citep{zhao2015, mariescu2017}. These approaches improve computational efficiency by eliminating the need for iterative clustering and pairwise distance calculations \citep{zhao2015}. More importantly, incorporating stay time as a primary metric represents a conceptual shift away from density-based inference. Stay time captures the duration an individual remains in a location, providing a more behaviorally meaningful indicator of activity than raw point counts, which can be influenced by device sampling frequency or movement patterns \citep{zheng2010, gong2015}. This emphasis on temporal duration is supported by a growing body of work that integrates stay-point detection and temporal constraints into mobility analysis, highlighting the importance of duration for improving accuracy and interpretability \citep{wang2023}.  In this study, we compare our proposed algorithm (GHOST) with five state-of-the-art home detection algorithms: All-time clustering (A1), Stay-point (A2), DBSCAN, K-Means++, and SciKit-mobility home detection (SciKit).

\subsection{All-time clustering}

The All-time clustering method (A1) is a simple baseline that infers home primarily on spatial density patterns rather than night-by-night variation \citep{verma2024}. In practice, A1 aggregates all nighttime pings across the entire dataset and applies mean-shift clustering \citep{Cheng1995MeanShift} with a flat kernel (e.g. 250 m), where all points within the kernel radius contribute equally to local density estimates, and points outside the radius have no influence. Mean-shift clustering iteratively shifts points toward regions of higher local density until convergence, allowing clusters to emerge directly from the data and the algorithm adaptively determine their initial locations. This makes A1 flexible for irregular urban environments and well suited for datasets where the number of frequently visited locations varies across individuals. Similar density-based, all-time approaches have been used in prior mobility studies to infer home and other activity locations from mobile phone data, including work that applies mean-shift clustering to detect stay places and dominant activity locations from sparse cellular network records \citep{Kanasugi2013ATC, Yabe2020ATC}. However, A1 is highly sensitive to the choice of kernel bandwidth, which controls the balance between spatial detail and over-smoothing. If the bandwidth is too small, GPS noise can split a single residence into multiple clusters; if it is too large, nearby but distinct locations may be merged \citep{Kanasugi2013ATC}. Because the method relies on cumulative spatial density, it can also perform poorly for users with multiple residences, extended absences, or irregular schedules.

\subsection{Stay-point method}

The Stay-point method (A2) extends earlier stay detection frameworks by combining spatiotemporal filtering with hierarchical clustering to infer meaningful activity locations \citep{sadeghinasr2019a2, verma2024}. Following the approach introduced by \cite{li2008staypoint}, the algorithm first identifies stay points, defined as locations where a user remains within a specified spatial radius (e.g., 250 m) for at least a minimum duration (often set to longer thresholds at night, such as three hours, and/or a longer cumulative threshold such as 24 hours total). These stay points are then grouped using hierarchical clustering with a maximum intra-cluster distance (e.g., 250 m) to form stay regions representing frequently visited locations, and home is inferred as the most frequently visited stay region during the designated nighttime window (e.g., 8 PM–5 AM). A key advantage of A2 is that it filters raw trajectories into behaviorally meaningful dwell events before clustering, which reduces sensitivity to transient movements, GPS noise, and irregular sampling \citep{li2008staypoint}. By incorporating dwell time, A2 is generally more robust than purely spatial clustering when data are noisy or sparse, and has therefore been widely adopted to extract significant locations such as home and work from GPS trajectories and related location histories \citep{li2008staypoint, verma2024, sadeghinasr2019a2}. However, A2’s performance depends strongly on the choice of spatial and temporal thresholds, which may not generalize across different urban forms, sampling rates, or behavioral contexts \citep{verma2024}. Thresholds that work well in dense urban settings can fail in suburban or rural areas, or for users with irregular schedules, frequent nighttime mobility, or short home dwell periods. In addition, like other clustering-based approaches, A2 can merge nearby but distinct locations when the clustering distance is too permissive, limiting its reliability in complex activity environments \citep{verma2024}.


\subsection{DBSCAN}

Among the most commonly used approaches are density-based algorithms, particularly Density-Based Spatial Clustering of Applications and Noise (DBSCAN) \citep{ester1996}. DBSCAN is frequently applied to nighttime GPS observations, where the densest spatial cluster is designated as an individual’s home location \citep{chen2017}. The algorithm groups spatially proximate points into clusters while treating low-density observations as noise, allowing the number of clusters to emerge directly from the data rather than being specified in advance. It is governed by two parameters: a neighborhood radius (\texttt{eps}) that defines spatial proximity and a minimum number of points (\texttt{MinPts}) required to form a cluster \citep{deng2020dbscan}. The minimum point requirement improves robustness by excluding sparse observations and emphasizing consistently visited locations instead of transient or noisy points. DBSCAN has proven effective in identifying dense concentrations of nighttime GPS points corresponding to residential locations, particularly due to its ability to capture irregularly shaped clusters and tolerate noisy or intermittently sampled data \citep{schubert2017}. As a result, it has been widely used for home and anchor-location inference across large-scale GPS trajectory datasets and geotagged social media data \citep{li2020mobilitypattern, mourtakos2024reconstructing, jurdak2015twitter, bora2014mobility}. However, its performance is highly sensitive to the choice of \texttt{eps} and \texttt{MinPts}, which are often difficult to determine and may not generalize across different spatial contexts or data densities. For example, parameter settings that perform well in dense urban environments may fail in suburban or rural areas, and the algorithm may struggle when clusters exhibit varying densities, potentially merging distinct locations or overlooking smaller but meaningful clusters \citep{gupta2023dbscan, ahmad2015performance}.


\subsection{K-Means++}

To address some of these challenges, centroid-based clustering approaches like K-MEANS++ have been explored for home detection tasks, offering a different methodological perspective rooted in partitioning data into a fixed number of clusters based on proximity to calculated centroids. Centroid-based clustering methods like K-Means++ are commonly applied in activity detection studies due to their computational efficiency, ease of interpretation, and widespread implementation. Prior work has applied K-Means++ to cluster individual mobility trajectories and activity patterns derived from mobile phone GPS data, using the resulting clusters to infer dominant anchor places such as home and work \citep{hao2024anchorplaces, yin2021activitytravel}. Similarly, clustering-based analyses of urban activity patterns have used K-Means++ alongside other methods to identify stable spatial concentrations of human activity that often correspond to residential or frequently visited locations \citep{bektemyssova2025spatial}. K-Means++ improves upon the standard K-Means algorithm by introducing a more informed initialization strategy for selecting initial centroids. Rather than choosing all centroids at random, K-Means++ selects the first centroid randomly and then chooses each subsequent centroid with a probability proportional to the squared distance from the nearest existing centroid, which helps reduce poor convergence and improves clustering stability \citep{arthur2007kmeans++}. However, K-Means++ assumes clusters are roughly convex and similar in size, limiting its effectiveness for irregular or noisy mobility traces \citep{arthur2007kmeans++}. In addition, the method requires specifying the number of clusters (k) in advance and is sensitive to outliers—both of which pose challenges when working with real-world GPS data where the number and structure of frequently visited locations can vary substantially across individuals \citep{bahmani2012}.


\subsection{SciKit-Mobility Home Detection}

The SciKit-mobility home detection algorithm (SciKit) is part of an open-source Python framework designed to support standardized and reproducible analyses of human mobility data \citep{pappalardo2022scikit}. The algorithm (e.g., \texttt{home\_location()}) infers user residences by filtering nighttime location points and selecting the most frequently visited coordinate as the home location. Its only key parameter is the nighttime window (start and end time), which can be adjusted for different sampling rates and temporal resolutions. The package provides standardized implementations of commonly used mobility measures, enabling more direct comparison across datasets and applications. As a result, the SciKit mobility package has been widely adopted in studies that rely on consistent and reproducible home inference. For example, \cite{rayat2023twitter} used it to estimate home locations from geotagged Twitter data; \cite{RoblesCruz2024Scikit} applied it to GPS trajectories to detect stay regions in health-related mobility research; \cite{Blake2025Scikit} employed its built-in functions (e.g., \texttt{max\_distance\_from\_home()} and \texttt{number\_of\_locations()}) to quantify individual mobility ranges; and \cite{Girolami2021Scikit} used it to identify dense residential clustering and individual mobility behavior from smartphone data. These applications demonstrate the flexibility of the SciKit-mobility framework across diverse spatiotemporal data sources; however, although the package is widely used to compute mobility metrics, relatively few studies rely on its home detection algorithm as the primary method for residential inference. In many cases, home locations serve as intermediate inputs rather than a focal methodological component, and key assumptions, such as reliance on nighttime presence and a single dominant home location, are rarely examined in detail. The approach may also struggle with multi-residence behavior, irregular schedules, or sparse nighttime observations, and the limited sensitivity analysis of parameter choices in prior work further motivates the need for systematic evaluation of SciKit’s home detection performance. 

Table \ref{tab:citations} summarizes the sources that have used these algorithms and the type of environment they were used in. 

\begin{table}[H]
\centering
\caption{Summary of representative home detection algorithms, their literature sources, dataset types, and practical applications.}
\label{tab:citations}
\footnotesize
\setlength{\tabcolsep}{4pt}
\renewcommand{\arraystretch}{1.2}

\begin{tabularx}{\textwidth}{
>{\raggedright\arraybackslash}p{2.2cm}
>{\raggedright\arraybackslash}X
>{\raggedright\arraybackslash}X
>{\raggedright\arraybackslash}X
}
\toprule
Algorithm & Sources & Dataset Types Used & Application \\
\midrule

A1  &
\cite{Yabe2020ATC, Cheng1995MeanShift, verma2024, Kanasugi2013ATC} &
GPS trajectories; mobile phone records; sparse cellular traces &
Baseline residential inference using cumulative nighttime density; suitable for detecting dominant locations when sampling is dense and spatially stable \\

A2 &
\cite{sadeghinasr2019a2, verma2024, li2008staypoint} &
GPS trajectories; smartphone logs; activity tracking datasets &
Extraction of meaningful dwell-based activity locations; supports behavior-aware home inference and filtering of transient movement noise \\

DBSCAN &
\cite{chen2017, li2020mobilitypattern, mourtakos2024reconstructing, jurdak2015twitter, bora2014mobility} &
GPS trajectories; social media geotags; mobile phone mobility datasets &
Density-based clustering for robust home detection in noisy or irregular datasets; effective for isolating stable nighttime activity clusters \\

K-Means++ &
\cite{hao2024anchorplaces, yin2021activitytravel, bektemyssova2025spatial} &
GPS mobility traces; smartphone location histories; urban activity datasets &
Computationally efficient partitioning of mobility traces into major activity centers; used when cluster structure is relatively stable and predefined \\

SciKit &
\cite{pappalardo2022scikit, rayat2023twitter, RoblesCruz2024Scikit, Blake2025Scikit, Girolami2021Scikit} &
GPS trajectories; social media mobility data; smartphone-derived datasets &
Reproducible home detection integrated into mobility analytics pipelines; supports consistent cross-study comparisons and metric computation \\

\bottomrule
\end{tabularx}
\end{table}


\section{Methodology}

\subsection{Overview of GHOST}

\begin{figure}[ht!]
    \centering
    
\includegraphics[width=\linewidth]{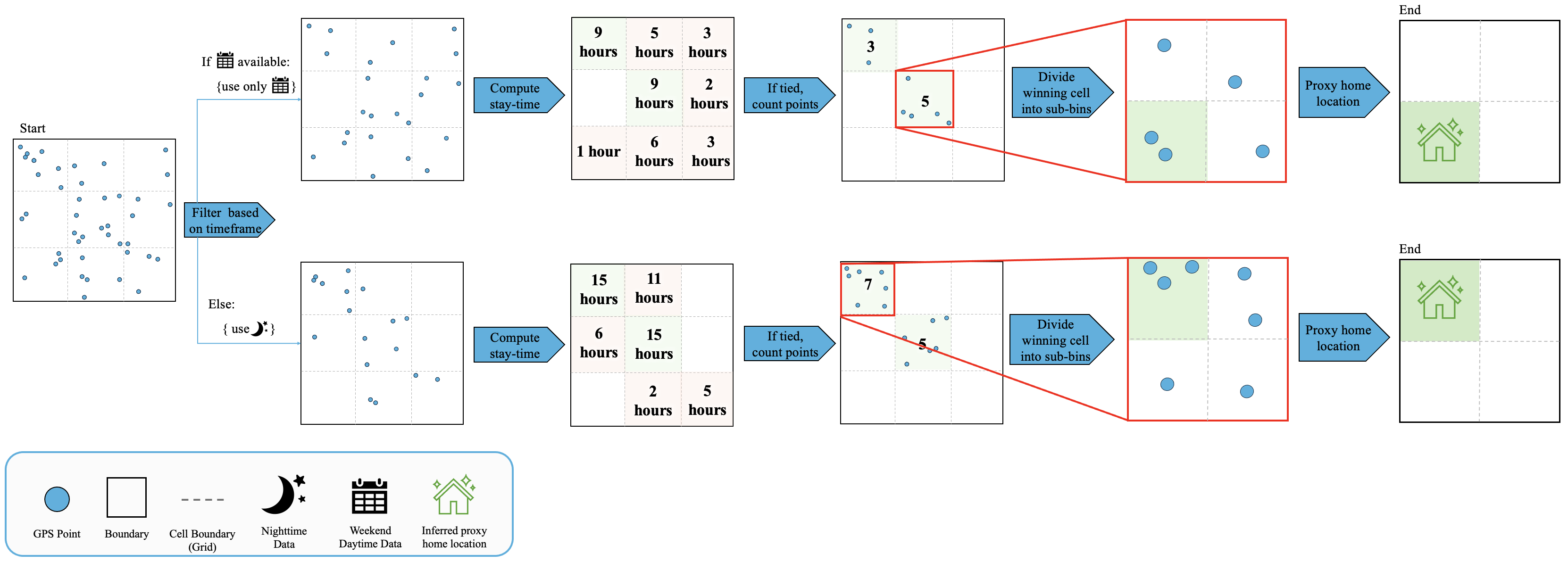}
    \caption{Home Location Inference Algorithms}
    \label{fig:ghost_flow}
\end{figure}

The core contribution of this work is the GHOST algorithm, which identifies a user’s home by selecting the spatial grid cell that exhibits the longest stay-time during nighttime hours, with an adaptive weekend fallback to ensure robustness when nighttime data are sparse. A schematic overview of the workflow is provided in Figure~\ref{fig:ghost_flow}, and the full procedure is detailed in Algorithm~\ref{alg:ghost}.

Prior research has demonstrated that home locations can often be inferred by identifying frequently visited locations during nighttime periods, under the assumption that users are typically at home at night \citep{calabrese2011estimating, saxon2021empirical}. However, methods relying exclusively on nighttime clustering can fail when users generate limited nighttime data due to irregular schedules, data sparsity, or device-level constraints. To address this limitation, GHOST incorporates a built-in weekend fallback mechanism. When no nighttime GPS points are available for a user, the algorithm shifts to identifying the most consistently occupied location during daytime hours on weekends (e.g., Saturday–Sunday, 8:00 AM–8:00 PM). This dual-strategy broadens the applicability of the method across diverse mobility behaviors while retaining otherwise discarded observations that may still contain meaningful residential signals. We distinguish these as \textit{night-inferred} and \textit{weekend-inferred} homes, respectively, to ensure methodological transparency.

A defining strength of GHOST is its avoidance of computationally intensive clustering operations. Rather than relying on pairwise distance calculations, density estimation, or iterative centroid optimization, as required by methods described in this study, GHOST employs a simple grid-based representation of space, and evaluates the stay-time occurring in each cell. 

The process begins by projecting raw GPS points for a resident into a local metric coordinate system (e.g., UTM) to enable accurate spatial calculations. The study area is then divided into square grid cells of a user-defined size ($X \times X$ meters), a customizable hyperparameter that balances spatial precision with data aggregation. A key innovation of GHOST lies in its home cell selection criterion. Many existing grid-based approaches select the cell with the highest number of GPS points \citep{zhao2022estimating, zhang2023situational}, an approach that can be misleading when point density reflects brief bursts of activity or noisy sampling, rather than a consistent pattern of residence. Instead, GHOST evaluates each grid cell using a \textit{stay-time} metric, defined as the temporal span between the first and last GPS observations recorded within that cell. The cell where the resident accumulates the longest stay-time is selected as the primary candidate for their proxy home location. By capturing the duration of presence rather than point frequency, stay-time provides a more robust indicator of consistent residence and directly reflects the reproducible nature of human mobility patterns \citep{gonzalez2008understanding}. 

In cases where multiple cells exhibit an equal maximum stay-time, the algorithm employs a robust hierarchical tie-breaking system: (1) the cell with the greater number of unique nights (or unique weekend days for fallback) is preferred, (2) if a tie persists, the cell with the highest total GPS point count serves as the final tie-breaker. Once the winning grid cell is identified, GHOST performs an adaptive intra-cell refinement step to improve spatial precision. The selected cell is subdivided into equal-area bins sized as the larger of 3 meters or one-tenth of the grid resolution. Point density is evaluated within each bin, and the home location is inferred as the centroid of GPS observations contained in the densest bin. If binning is infeasible due to sparse observations (e.g., fewer than three points), the algorithm falls back to the mean coordinates of all points in the winning cell, and ultimately to the original grid cell center. This refinement preserves computational efficiency while improving robustness to GPS noise, sampling variability, and local clustering artifacts. Finally, the inferred home coordinates are transformed back to geographic space (latitude/longitude) and reported along with supporting metrics, specifically outputting the exact spatial refinement method applied (e.g., densest bin, cell mean, or grid centroid) to ensure maximum methodological transparency and confidence. The temporal filtering windows are fully configurable, with default nighttime hours of 10 PM to 6 AM and weekend fallback hours of 8 AM to 8 PM on Saturdays and Sundays. These defaults can be adjusted to accommodate different cultural contexts, work schedules, or behavioral patterns. 

\begin{algorithm}[H]
\footnotesize
\caption{GHOST: Grid-based Home detection via Stay-Time Algorithm}\label{alg:ghost}
\KwIn{GPS dataset $D = \{p\_1, p\_2, \ldots, p\_n\}$ where each point $p\_i = (user\_id, timestamp, lat, lon)$}
\KwIn{Grid size $g$ (meters), night window $[t\_{start}, t\_{end}]$, weekend window $[t\_{weekend\_start}, t\_{weekend\_end}]$}
\KwOut{Home location $(lat\_{home}, lon\_{home})$ for each user}

\textbf{Data Preprocessing}\;
$D\_{geo} \leftarrow$ Convert $D$ to GeoDataFrame with CRS: EPSG:4326\;
$(D\_x, D\_y) \leftarrow$ Project coordinates from EPSG:4326 to local UTM CRS\;
$D\_{time} \leftarrow$ Extract hour and day-of-week features from timestamps\;

\textbf{Temporal Filtering}\;
$D\_{night} \leftarrow \{p\_i \in D : hour(p\_i.timestamp) \in [t\_{start}, t\_{end}] \cup [0, t\_{end}]\}$\;
$D\_{weekend} \leftarrow \{p\_i \in D : dayofweek(p\_i.timestamp) \in \{5,6\} \land hour(p\_i.timestamp) \in [t\_{weekend\_start}, t\_{weekend\_end}]\}$\;

\textbf{Home Detection}\;
\For{each user $u$ in unique users}{
    $D\_u \leftarrow$ Filter $D$ by $user\_id = u$\;
    $D\_{u\_night} \leftarrow$ Filter $D\_{night}$ by $user\_id = u$\;
    
    \If{$|D\_{u\_night}| > 0$}{
        $inference\_source \leftarrow$ 'night'\;
        $D\_{filtered} \leftarrow D\_{u\_night}$\;
    }
    \Else{
        $inference\_source \leftarrow$ 'weekend'\;
        $D\_{u\_weekend} \leftarrow$ Filter $D\_{weekend}$ by $user\_id = u$\;
        $D\_{filtered} \leftarrow D\_{u\_weekend}$\;
    }
    
    \If{$|D\_{filtered}| = 0$}{
        Return $(null, null)$ for user $u$\;
        Continue to next user\;
    }
    
    \textbf{Grid Assignment}\;
    $D\_{grid} \leftarrow$ Assign each point to grid cell: $cell\_x = round(x / g) \times g$, $cell\_y = round(y / g) \times g$;
    
    \textbf{Stay-Time Calculation}\;
    \For{each unique grid cell $c$ in $D\_{grid}$}{
        $points\_in\_cell \leftarrow$ Filter $D\_{grid}$ by $cell = c$\;
        $unique\_nights \leftarrow |unique(dates(points\_in\_cell))|$\;
        $total\_points \leftarrow |points\_in\_cell|$\;
        $stay\_time \leftarrow max(timestamps(points\_in\_cell)) - min(timestamps(points\_in\_cell))$\;
        $cell\_stats[c] \leftarrow (unique\_nights, total\_points, stay\_time)$\;
    }
    
    \textbf{Home Cell Selection}\;
    $home\_cell \leftarrow$ Select cell with maximum $stay\_time$\;
    If tie: select cell with maximum $unique\_nights$\;
    If tie: select cell with maximum $total\_points$\;
    
    \textbf{Intra-Cell Refinement \& Coordinate Transformation}\;
    $bin\_size \leftarrow \max(3.0, g / 10.0)$\;
    $P_{home} \leftarrow$ points in $home\_cell$\;
    \If{$|P_{home}| < 3$}{
        $(x\_{home}, y\_{home}) \leftarrow mean(P_{home})$\;
        $method \leftarrow$ 'mean\_cell\_points'\;
    }
    \Else{
        Divide $home\_cell$ into $bin\_size \times bin\_size$ equal-area sub-bins\;
        $best\_bin \leftarrow$ sub-bin containing the maximum number of points\;
        $(x\_{home}, y\_{home}) \leftarrow mean(points \in best\_bin)$\;
        $method \leftarrow$ 'densest\_bin\_centroid'\;
    }
    \If{refinement fails (NaN)}{
        $(x\_{home}, y\_{home}) \leftarrow$ Centroid of $home\_cell$\;
        $method \leftarrow$ 'grid\_centroid'\;
    }
    $(lat\_{home}, lon\_{home}) \leftarrow$ Transform $(x\_{home}, y\_{home})$ back to EPSG:4326\;

    Return $(lat\_{home}, lon\_{home}, inference\_source, cell\_stats[home\_cell])$\;
}

\end{algorithm}

\subsection{Algorithm Robustness and Advantages}

The GHOST algorithm incorporates several design decisions that make it particularly robust and advantageous for real-world applications:

\subsubsection{Computational Efficiency} Unlike traditional clustering methods that require pairwise distance calculations or iterative optimization, GHOST operates through simple spatial binning and aggregation operations. The algorithm's time complexity is linear with respect to the number of GPS points, making it highly scalable for large datasets. The grid-based approach eliminates the need for complex geometric computations, while the hierarchical selection criterion requires only basic arithmetic operations.

\subsubsection{Noise Resistance}
The algorithm’s robustness to GPS noise and outliers stems from its grid-based spatial binning approach. Individual GPS inaccuracies and the "pull" of multi-path GPS errors are absorbed within grid cells. Meanwhile, aggregated metrics such as stay-time and unique nights further mitigate the effects of sporadic data errors. GHOST then employs an adaptive densest sub-bin centroid strategy, dynamically subdividing the cell with the most stay-time into equal-area micro-bins (e.g., 2–3 meters). It defines the home location as the centroid of GPS points exclusively within the densest bin, filtering out spatial scatter to pinpoint the true physical structure and reduce sensitivity to clustering artifacts and sampling irregularities. If binning is infeasible, fallback strategies ensure a stable estimate, prioritizing consistent spatial patterns over random noise.

\subsubsection{Adaptive Temporal Filtering} The weekend fallback mechanism significantly increases the algorithm's applicability by accommodating users with diverse behavioral patterns. This feature is particularly valuable for populations that may not generate sufficient nighttime GPS data due to work schedules, lifestyle choices, or technical limitations.

\subsubsection{Configurable Parameters} The algorithm's modular design allows researchers to fine-tune parameters based on specific study contexts, including grid, temporal windows, and coordinate projection systems. This flexibility enables adaptation to different geographic regions, population characteristics, and research objectives.

\subsection{Data Requirements}
The algorithm requires GPS data with the following specifications: (1) Format: GPX files, CSV files, or directories containing multiple files; (2) Required Columns: user\_id, timestamp (datetime), latitude, longitude; (3) Coordinate Reference System: WGS84 (EPSG:4326) for input data; (4) Temporal Coverage: Minimum 7 days recommended for reliable inference; (5) Spatial Accuracy: GPS accuracy within 10-20 meters preferred; and (6) Data Quality: Handles missing data and temporal gaps automatically. 

\subsection{Software Development}
The \texttt{ghost} package is structured for modularity, usability, and extensibility which enables it to serve as both a ready-to-use tool and a platform for future research: \url{https://github.com/SERMOS-LAB/Grid-Based-Home-Detection}. 

\subsubsection{Architecture and Design}
The software is organized into distinct modules for I/O, preprocessing, algorithms, validation, and command-line interface (CLI) logic. The core of the programmatic interface is the \texttt{HomeDetector} class, which encapsulates the entire workflow. This class-based design provides a clean, high-level API for users, abstracting away the underlying complexity. The \texttt{HomeDetector} class provides a unified workflow that handles both single-user and bA1h processing scenarios automatically. Users can process individual GPX files or entire directories containing multiple users' data with the same simple interface, making the tool highly scalable for large-scale studies.

Configuration management is a central design feature that supports multiple use cases. All parameters, from file paths and algorithm settings (e.g., \texttt{grid\_size}) to temporal filters, can be specified via a YAML configuration file for reproducible research workflows, passed directly as keyword arguments for interactive analysis, or mixed together with keyword arguments overriding configuration file settings. This flexibility enables both reproducible script-based workflows and dynamic parameter exploration in environments like Jupyter notebooks.

\subsubsection{Key Components and Functionality}
The functionality of the \texttt{ghost} package is summarized in Table \ref{tab:software_components}. The software is designed for bA1h and single-user processing. When provided with a directory of GPX files or a CSV with a user ID column, it automatically iterates through each user, runs the detection pipeline, and aggregates the results into a single output file.

\begin{table}[h!]
\centering
\caption{Key Components of the \texttt{ghost} Software Package}
\label{tab:software_components}
\begin{tabular}{ll}
\hline
\textbf{Component} & \textbf{Description and Purpose} \\ \hline
\texttt{HomeDetector} Class & High-level API for managing the end-to-end workflow. \\
& Initializes with default, config file, or keyword parameters. \\
\\
I/O Module & Handles data ingestion from various formats (GPX, CSV). \\
& Standardizes data into a GeoPandas GeoDataFrame. \\
\\
Preprocessing Module & Contains functions for coordinate projection and temporal \\
& feature extraction (e.g., hour of day). \\
\\
Algorithms Module & Implements the core grid-based detection logic and is \\
& designed to accommodate other algorithms (e.g., DBSCAN). \\
\\
Validation Module & Provides tools to compare inferred home locations against \\
& ground-truth data and compute accuracy metrics (e.g., haversine distance). \\
\\
Command-Line Interface & A Typer-based CLI for running the entire pipeline, \\
(CLI) & including detection, validation, and plotting, from the terminal. \\
\\
BA1h Processing & Automatically handles multiple users from GPX directories \\
& or CSV files with user ID columns. \\
\\
Plotting Module & Generates static and interactive visualizations of GPS traces \\
& and inferred home locations for individual users or bA1hes. \\
\\
Documentation & Comprehensive API documentation, tutorials, and example \\
& workflows for both CLI and programmatic usage. \\
\hline
\end{tabular}
\end{table}

\subsubsection{Unified Workflow}
The \texttt{ghost} package implements a unified workflow that automatically adapts to different input scenarios. For single-user analysis, the \texttt{HomeDetector} class processes individual GPX files and returns detailed results with confidence metrics. For bA1h processing, the same interface automatically detects multiple users from directory structures or CSV files, processes each user independently, and aggregates results into a comprehensive output file. This design eliminates the need for users to write custom loops or data management code, significantly reducing the barrier to entry for large-scale mobility studies.

\subsubsection{Reproducibility and Extensibility}
Reproducibility is ensured by a flexible configuration management system and the project's open-source availability, which allows for transparent inspection and replication of any analysis. The modular architecture facilitates extensibility. For instance, a novel detection algorithm can be readily integrated within the existing framework. This enables direct, systematic comparisons against the baseline grid-based method under identical data handling and validation conditions. To support its reliability and accessibility for the research community, the package is distributed with a comprehensive suite of unit tests, detailed documentation, and practical example workflows.

\subsubsection{Validation and Reproducibility}
The package includes comprehensive validation tools that enable direct comparison of inferred home locations against ground truth data. The validation module provides standard accuracy metrics including mean error, median error, and percentage of predictions within specified distance thresholds. Additionally, the package supports reproducible research through version-controlled configuration files, comprehensive unit tests, and detailed logging of all processing steps. All algorithms, preprocessing steps, and validation metrics are fully documented and open-source, enabling transparent replication and extension of research findings.

\section{The Data}

To evaluate the performance of the proposed home detection algorithms, we use two complementary datasets that capture both idealized, high-quality mobility data and controlled, ground-truth conditions. We begin with the BostonWalks dataset from the Massachusetts Institute of Technology (MIT), which includes GPS trajectories from 377 users with self-reported home locations \citep{meister2025bostonwalks}. Collected through smartphone-based travel diaries, this dataset offers high-resolution, consistently sampled trajectories with rich behavioral context, making it well suited for assessing performance under ideal data conditions. We then complement this analysis with a GeoTracker dataset collected from volunteer participants who recorded GPS trajectories for at least seven consecutive days across Florida, Colorado, and Mississippi. The participants then self-reported their true home locations to us directly. While this dataset includes verified home locations, it reflects common real-world challenges such as incomplete temporal coverage, irregular sampling, and behavioral variability. Together, these datasets provide a comprehensive evaluation framework that assesses both algorithm accuracy under optimal conditions and robustness under sparse, imperfect mobility data scenarios.

\subsection{BostonWalks Dataset}

To evaluate algorithm performance under highly favorable data conditions, we utilize the BostonWalks dataset provided by MIT \citep{meister2025bostonwalks}, which represents a large-scale, high-quality mobility dataset. The dataset includes 990 participants from the Boston Metropolitan Area, with a median tracking duration of 23 days, approximately 235,000 recorded tracks, and 172,000 identified stay events. Unlike typical large-scale mobile location datasets obtained from commercial vendors, BostonWalks is constructed using smartphone-based travel diaries, resulting in exceptionally rich and structured data.

The dataset provides a high level of detail and enrichment, including precise trajectory waypoints, comprehensive sociodemographic indicators, and state-of-the-art positional accuracy. Additionally, the dataset is pre-processed to include derived information such as transportation mode detection (e.g., car, walking, biking, rail transit, and airplane), spatial and temporal movement patterns, and trip purpose classification. These features make BostonWalks an ideal dataset for algorithm evaluation, as it minimizes common data limitations such as missingness, irregular sampling, and behavioral ambiguity.

From this dataset, 377 participants voluntarily labeled their stay points as ``Home,'' and only these users were retained for validation. The resulting subset contains nearly 30 million GPS points, providing a dense and information-rich environment for testing. Due to its structured nature and extensive enrichment, BostonWalks represents an idealized data scenario that is not fully representative of vendor-based mobility datasets, which often suffer from sparsity, noise, and incomplete coverage. As such, this dataset is primarily used to evaluate the scalability and robustness of home detection algorithms under conditions of highly noisy data.

\subsection{GeoTracker Dataset}

To complement the BostonWalks dataset, we use the GeoTracker dataset, which was collected through a controlled data collection effort involving ten voluntary participants whom we recruited directly. Participants were located across Florida, Colorado, and Mississippi and were instructed to record their GPS data continuously for at least seven consecutive days, with flexibility in timing to capture natural variability in behavior. On average, users collected approximately 12 days of data, resulting in roughly 200,000 total GPS points. Daily temporal coverage averaged approximately 48\% over a 24-hour period, reflecting realistic inconsistencies such as signal loss, device inactivity, and irregular usage patterns.

A key advantage of this dataset is the availability of verified ground truth. Each participant independently provided their true home location coordinates, enabling direct and transparent accuracy evaluation. This addresses a major limitation in many mobility studies, where true home locations are unknown or inferred indirectly. Due to its shorter collection period (approximately 1–2 weeks), reliance on a single application, and smaller user base, the GeoTracker dataset represents a relatively sparse, incomplete, and behaviorally realistic data environment.

In contrast to BostonWalks, GeoTracker more closely reflects the types of limitations encountered in real-world mobile location data, including missing data, inconsistent sampling, and reduced temporal coverage. As such, it serves as a critical benchmark for evaluating the accuracy of home detection algorithms under controlled and less noisy data conditions. Table~\ref{tab:user_gps_summary} in Appendix~\ref{app:geotracker} summarizes GPS activity for all GeoTracker users, including total points, duration, daily statistics, and coverage (the proportion of theoretical recording hours with captured data), offering insight into the temporal completeness of each user's dataset.

\section{Sensitivity Analysis}

\begin{figure}[H]
    \centering
    \includegraphics[width=1.0\linewidth]{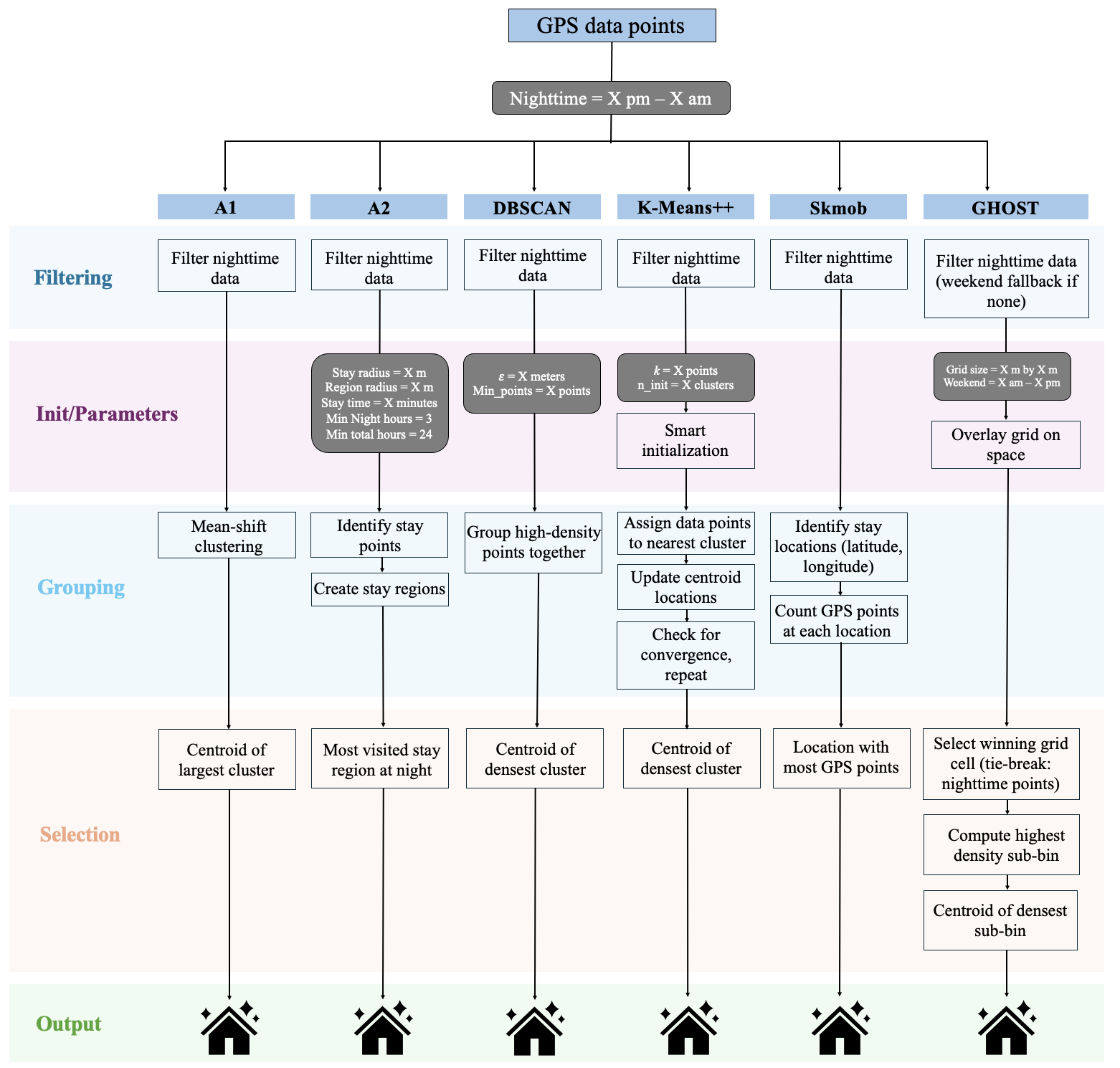}
    \caption{Home Location Inference Algorithms}
    \label{fig:algo_figure}
\end{figure}

To evaluate the robustness of home detection methods, we conducted a sensitivity analysis across six algorithms: A1, A2, DBSCAN, K-Means++, SciKit, and GHOST. A1 and A2 \citep{sadeghinasr2019a2} were selected for their conceptual simplicity and relevance to the core principles underlying GHOST. While A1 represents a straightforward clustering-based approach, A2 incorporates stay-time and dwell-time thresholds to identify meaningful activity locations, making it more directly comparable to GHOST. However, unlike GHOST’s grid-based framework, A2 still relies on clustering procedures, enabling comparison between clustering- and grid-based approaches while both consider temporal duration. DBSCAN and K-Means++ represent widely used density-based and centroid-based clustering methods for identifying activity hotspots \citep{li2022kmeans, ran2021kmeans, yoo2020dbscan, kato2024dbscan, yang2022dbscan}, while SciKit-mobility was included due to the widespread adoption of the package within the Python mobility analysis community \citep{pappalardo2022scikit, Blake2025Scikit, Girolami2021Scikit, rayat2023twitter, RoblesCruz2024Scikit}. These similarities and differences in algorithmic logic are presented in Figure \ref{fig:algo_figure}. 

To assess algorithm robustness and scalability, the BostonWalks users were divided into an 80\% training set and a 20\% testing set using a fixed random seed (\texttt{global\_seed} = 42), while the GeoTracker dataset was treated as an external testing set. All parameter tuning was performed exclusively on the BostonWalks training set to avoid information leakage and ensure a fair evaluation of out-of-sample performance. Across all algorithms, temporal filtering was varied systematically by testing nighttime windows spanning start hours from 20:00 to 22:00, and end hours from 05:00 to 07:00. Additional method-specific parameters were also evaluated, including bandwidth for A1, stay distance and dwell-time thresholds for A2, epsilon and minimum points for DBSCAN, cluster number and initialization settings for K-MEANS++, and grid size for GHOST. For algorithms involving random initialization, multiple \texttt{random\_state} values were tested to evaluate convergence stability and reproducibility.

For each algorithm, all parameter combinations were first evaluated on the BostonWalks training set, and the configuration minimizing distance to the ground-truth home locations was selected as the optimal setting. These optimized parameters were then applied without modification to both the held-out BostonWalks testing set and the GeoTracker dataset, enabling evaluation of both parameter sensitivity and cross-dataset generalizability under differing mobility data conditions. Each algorithms' unique parameter combinations are described below.

Performance was then evaluated using Mean Absolute Error (MAE) and Root Mean Square Error (RMSE), which quantify the spatial discrepancy between inferred and ground-truth home locations. MAE represents the average distance error across all users, providing a measure of how far predictions are from true home locations on average. In contrast, RMSE places more weight on larger errors, making it more sensitive to cases where predicted locations are far from the true location. These metrics provide a balanced view of overall accuracy and sensitivity to outliers, which is especially important in home detection tasks. Let $d_i$ denote the geographic distance (in meters) between the predicted and true home location for user $i$, computed using the haversine formula:

\begin{center}
$d_i = \text{dist}\left(
(\mathrm{lat}_i^{\mathrm{pred}}, \mathrm{lon}_i^{\mathrm{pred}}),
(\mathrm{lat}_i^{\mathrm{true}}, \mathrm{lon}_i^{\mathrm{true}})
\right)$
\end{center}

where $(\mathrm{lat}_i^{\mathrm{pred}}, \mathrm{lon}_i^{\mathrm{pred}})$ and $(\mathrm{lat}_i^{\mathrm{true}}, \mathrm{lon}_i^{\mathrm{true}})$ denote the inferred and ground-truth home coordinates, respectively, and $N$ is the total number of users.

The mean absolute error (MAE) is defined as the average of the individual distance errors, providing a measure of the typical magnitude of spatial deviation:

\begin{center}
$\mathrm{MAE} = \frac{1}{N}\sum_{i=1}^{N} d_i$
\end{center}

The root mean square error (RMSE) is defined as the square root of the mean of squared distance errors, placing greater emphasis on larger deviations and thus increasing sensitivity to outliers:

\begin{center}
$\mathrm{RMSE} = \sqrt{\frac{1}{N}\sum_{i=1}^{N} d_i^2}$
\end{center}

\subsubsection{All-time clustering (A1)}
A1 was implemented using mean-shift clustering on aggregated nighttime points. Sensitivity was evaluated across kernel bandwidths (\texttt{bandwidth\_m} = 20, 50, 150, 250 m). These parameters control the degree of spatial smoothing and data density required to identify a dominant home location and were selected to align with spatial scales tested in DBSCAN and GHOST.

\subsubsection{Stay-point method (A2)}
The Stay-point method was evaluated by varying the spatial radius used to identify stays (\texttt{stay\_dist\_m} = 20, 50, 150, 250 m), the minimum dwell time required to form a stay (\texttt{stay\_time\_min} = 10, 25, 50 minutes), and the clustering radius used to group stays into regions (\texttt{region\_radius\_m} = 20, 50, 150, 250 m). Regions were classified as home if they accumulated at least three hours of nighttime dwell time or 24 hours total dwell time, consistent with prior implementations \citep{li2008staypoint}. Home locations were defined as the centroid of the qualifying region.

\subsubsection{DBSCAN}
DBSCAN was evaluated across neighborhood radius values (\texttt{eps} = 20, 50, 150, 250 m) and minimum cluster size thresholds (\texttt{MinPts} = 2, 4, 6). These parameters control sensitivity to spatial density and clustering granularity, allowing assessment of performance under both sparse and dense mobility traces.

\subsubsection{K-Means++}
K-Means++ was evaluated by varying the number of clusters (\texttt{k} = 2, 4, 6), the initialization seed (\texttt{random\_state} = 42, 100, 2048), and the number of initialization runs (\texttt{n\_init} = 20). For each user, home was inferred as the centroid of the dominant cluster. These parameters jointly assess sensitivity to initialization and convergence behavior.

\subsubsection{SciKit-mobility Home Detection}
The SciKit-mobility home detection algorithm \citep{pappalardo2022scikit} was evaluated as a standardized, parameter-light baseline. The method infers home as the most frequently observed nighttime location (coordinate point). Sensitivity analysis was limited to variations in nighttime window definitions as the algorithm does not use spatial or dwell-time parameters.

\subsubsection{GHOST}
GHOST was evaluated across grid resolutions (\texttt{grid\_size} = 20, 50, 150, 250 m), which define the spatial binning used to accumulate stay-time. These values parallel the spatial scales tested in DBSCAN and A1, enabling direct comparison between grid-based and clustering-based representations. Just like cluster size, smaller grid sizes improve spatial precision but increase sensitivity to sparsity, while larger grids promote stability at the cost of spatial detail. Table \ref{tab:sensitivity_parameters} displays the different parameter combinations tested, and Table~\ref{tab:frozen_params} displays the final parameters applied to the BostonWalks and GeoTracker testing sets.

\begin{table}[H]
\centering
\caption{Parameter ranges evaluated during sensitivity analysis.}
\label{tab:sensitivity_parameters}

\footnotesize
\setlength{\tabcolsep}{4pt}
\renewcommand{\arraystretch}{1.15}

\begin{adjustbox}{max width=\columnwidth}
\begin{tabular}{lllll}
\toprule

\textbf{Algorithm} &
\textbf{Parameter} &
\textbf{Values} &
\textbf{Start} &
\textbf{End} \\

\midrule

A1
& \texttt{bandwidth\_m}
& 20, 50, 150, 250
& 20--22
& 05--07 \\

\midrule

\multirow{3}{*}{A2}
& \texttt{stay\_dist\_m}
& 20, 50, 150, 250
& \multirow{3}{*}{20--22}
& \multirow{3}{*}{05--07} \\
& \texttt{stay\_time\_min}
& 10, 25, 50
&& \\
& \texttt{region\_radius\_m}
& 20, 50, 150, 250
&& \\

\midrule

\multirow{2}{*}{DBSCAN}
& \texttt{eps}
& 20, 50, 150, 250
& \multirow{2}{*}{20--22}
& \multirow{2}{*}{05--07} \\
& \texttt{MinPts}
& 2, 4, 6
&& \\

\midrule

\multirow{3}{*}{K-Means++}
& \texttt{k}
& 2, 4, 6
& \multirow{3}{*}{20--22}
& \multirow{3}{*}{05--07} \\
& \texttt{random\_state}
& 42, 100, 2048
&& \\
& \texttt{n\_init}
& 20
&& \\

\midrule

SciKit
& --
& --
& 20--22
& 05--07 \\

\midrule

GHOST
& \texttt{grid\_size}
& 20, 50, 150, 250
& 20--22
& 05--07 \\

\bottomrule
\end{tabular}
\end{adjustbox}
\end{table}

\begin{table}[H]
\centering
\caption{Frozen parameter values applied to testing datasets.}
\label{tab:frozen_params}

\footnotesize
\setlength{\tabcolsep}{4pt}
\renewcommand{\arraystretch}{1.15}

\begin{adjustbox}{max width=\columnwidth}
\begin{tabular}{lllll}
\toprule

\textbf{Algorithm} &
\textbf{Parameter} &
\textbf{Value} &
\textbf{Start} &
\textbf{End} \\

\midrule

A1
& \texttt{bandwidth\_m}
& 20
& 22
& 05 \\

\midrule

\multirow{3}{*}{A2}
& \texttt{stay\_dist\_m}
& 50
& \multirow{3}{*}{20}
& \multirow{3}{*}{05} \\
& \texttt{stay\_time\_min}
& 10
&& \\
& \texttt{region\_radius\_m}
& 50
&& \\

\midrule

\multirow{2}{*}{DBSCAN}
& \texttt{eps}
& 20
& \multirow{2}{*}{21}
& \multirow{2}{*}{05} \\
& \texttt{MinPts}
& 4
&& \\

\midrule

\multirow{2}{*}{K-Means++}
& \texttt{k}
& 1
& \multirow{2}{*}{22}
& \multirow{2}{*}{05} \\
& \texttt{n\_init}
& 10
&& \\

\midrule

SciKit
& --
& --
& 20
& 05 \\

\midrule

GHOST
& \texttt{grid\_size}
& 50
& 22
& 06 \\

\bottomrule
\end{tabular}
\end{adjustbox}
\end{table}


\section{Results}

Using parameters optimized on the training set (Table~\ref{tab:frozen_params}), GHOST consistently outperformed all baseline methods across both the BostonWalks testing set and the GeoTracker dataset. This advantage was particularly pronounced in the GeoTracker dataset, where GHOST achieved minimal error, while maintaining strong performance in the more complex and noisy BostonWalks dataset. In contrast, clustering-based methods exhibited greater sensitivity to dataset characteristics, often resulting in higher errors and less stable performance.

\subsection{All-time clustering (A1)}

Starting with the first algorithm described in this study, A1 was implemented using a MeanShift bandwidth of 20 m and a nighttime window of 22:00–05:00. In the BostonWalks dataset, performance deteriorated substantially, with an MAE of 492,559.47 m and RMSE exceeding 1.73 million meters. In the GeoTracker dataset, A1 performed relatively well, achieving an MAE of 530.97 m and RMSE of 1,607.99 m. 

\subsection{Stay-point method (A2)}

The stay-point method was evaluated using a stay distance of 50 m, dwell time threshold of 10 minutes, region radius of 50 m, and a nighttime window of 20:00–05:00. In the BostonWalks dataset, A2 achieved improved performance, with an MAE of 14,045.74 m and RMSE of 113,601.40 m. In the GeoTracker dataset, A2 performed poorly overall, with an MAE of 27,102.29 m and RMSE of 76,244.58 m.  

\subsection{DBSCAN}

DBSCAN was evaluated using $\epsilon = 20$ m, $\texttt{min\_samples} = 4$, and a nighttime window of 21:00–05:00. In the BostonWalks dataset, DBSCAN performed relatively well compared to other clustering-based methods, achieving an MAE of 1,111.46 m and RMSE of 2,429.85 m. In the GeoTracker dataset, DBSCAN produced moderate performance, with an MAE of 4,248.53 m and RMSE of 8,296.29 m. 

\subsection{K-MEANS++}

K-MEANS++ was evaluated with $k = 1$ and a nighttime window of 22:00–05:00. In the BostonWalks dataset, performance remained weak, with an MAE of 269,897.57 m and RMSE exceeding 1.25 million meters. In the GeoTracker dataset, K-MEANS++ performed poorly, with an MAE of 19,467.40 m and RMSE of 37,942.42 m. Errors were driven by the centroid collapsing toward the mean of all points, which does not correspond to a meaningful activity location.

\subsection{SciKit-mobility Home Detection}

The SciKit-mobility method was evaluated using a nighttime window of 20:00–05:00. In the BostonWalks dataset, SciKit-mobility similarly exhibited instability, with an MAE of 194,213.07 m and RMSE of 1.33 million meters. In the GeoTracker dataset, SciKit-mobility produced highly variable results, with an MAE of 24,016.25 m and RMSE of 75,881.00 m. 

\subsection{GHOST}

Using a grid size of 50 m and a nighttime window of 22:00–06:00, GHOST achieved the best overall performance across both datasets. In the BostonWalks dataset, GHOST maintained strong performance, achieving an MAE of 786.53 m and RMSE of 1,938.65 m. In the GeoTracker dataset, GHOST produced extremely low errors, with an MAE of 22.33 m and RMSE of 35.30 m, indicating highly precise home detection across all users. While errors were larger relative to the GeoTracker dataset due to real-world noise and variability, GHOST still outperformed all competing algorithms. 

\subsection{Comparative Summary}

\begin{figure}[ht!]
    \centering
    
\includegraphics[width=\linewidth]{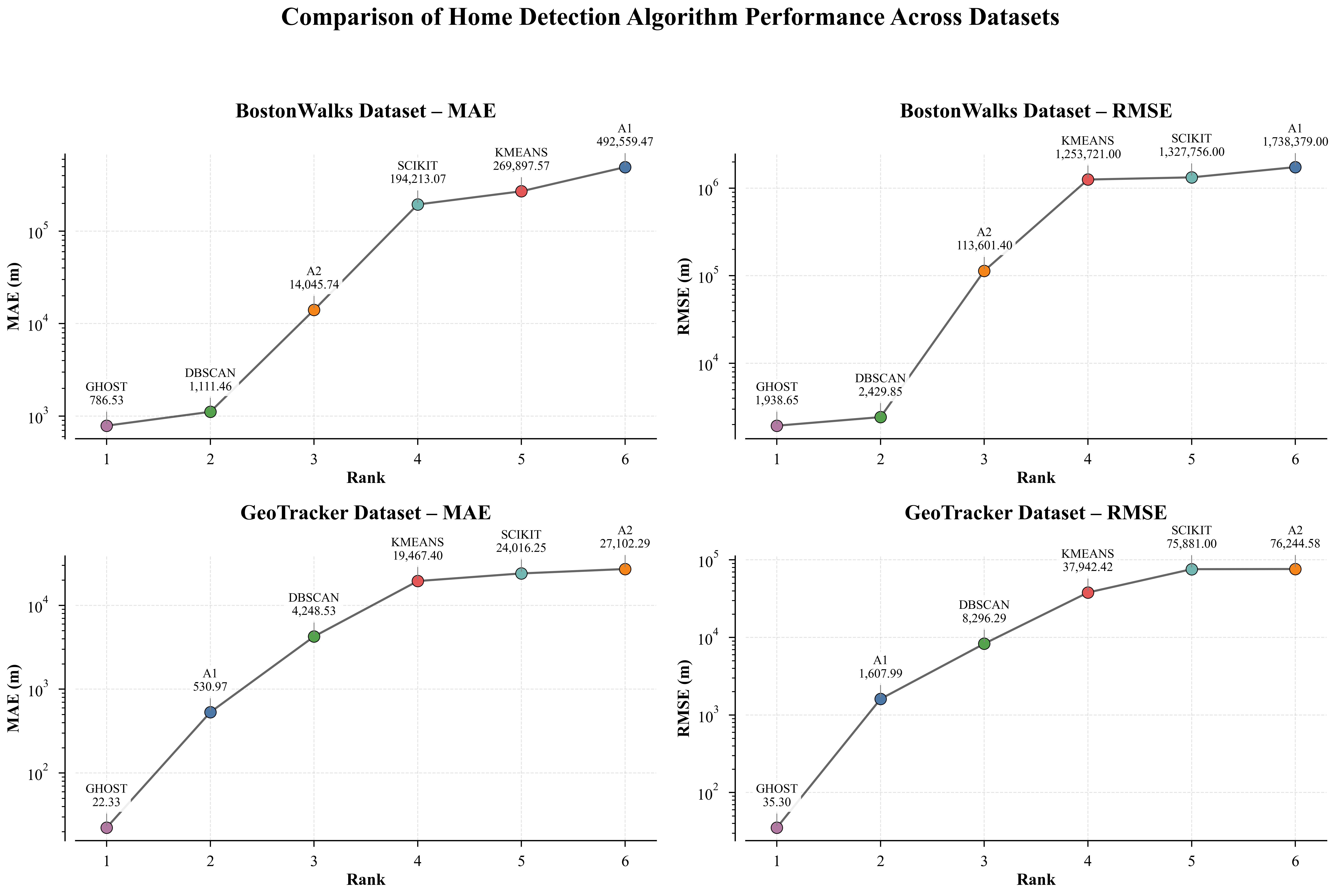}
    \caption{RMSE and MAE for Both Dataset}
    \label{fig:mae}
\end{figure}

Across both datasets, clear differences emerged between methodological approaches. Grid-based methods that incorporate stay time, as implemented in GHOST, consistently achieved the lowest errors and most stable performance. In contrast, clustering-based methods exhibited sensitivity to parameter selection, data sparsity, and irregular sampling, leading to large errors and reduced reliability. These results demonstrate that incorporating stay-time within a structured spatial framework provides a more robust and accurate approach to home location detection. Table~\ref{tab:performance_comparison}, along with Figure~\ref{fig:mae}, summarizes the MAE and RMSE performance of each algorithm on the BostonWalks and GeoTracker datasets.

\begin{table}[h!]
\centering
\caption{Home Detection Performance Across Datasets}
\label{tab:performance_comparison}
\begin{tabular}{lcccc}
\hline
\multirow{2}{*}{\textbf{Algorithm}} & \multicolumn{2}{c}{\textbf{BostonWalks}} & \multicolumn{2}{c}{\textbf{GeoTracker}} \\
\cline{2-5}
 & \textbf{MAE (m)} & \textbf{RMSE (m)} & \textbf{MAE (m)} & \textbf{RMSE (m)} \\
\hline
A1  & 492,559.47 & 1,730,000.00 & 530.97   & 1,607.99 \\
A2   & 14,045.74  & 113,601.40  & 27,102.29 & 76,244.58 \\
DBSCAN                  & 1,111.46   & 2,429.85    & 4,248.53  & 8,296.29 \\
K-MEANS++               & 269,897.57 & 1,250,000.00 & 19,467.40 & 37,942.42 \\
SciKit-mobility         & 194,213.07 & 1,330,000.00 & 24,016.25 & 75,881.00 \\
GHOST                   & 786.53     & 1,938.65     & 22.33     & 35.30 \\
\hline
\end{tabular}
\end{table}

\section{Discussion}

\subsection{Performance Across Datasets}

The results of this study demonstrate that home detection performance is strongly influenced by data quality, sampling consistency, and behavioral complexity. Across both datasets, GHOST consistently produced lower spatial error and more stable performance than competing approaches, highlighting the advantages of combining stay-time aggregation, adaptive temporal filtering, and a grid-based spatial framework.

The BostonWalks dataset highlights the challenges of applying home detection algorithms to noisy, irregular, and behaviorally complex mobility data. Across this dataset, many methods exhibited substantially large RMSE values, reflecting sensitivity to uneven sampling frequency and GPS noise. These effects were most pronounced for \textbf{K-MEANS++} and \textbf{A1}, which rely on centroid-based clustering approaches (i.e. mean-shift clustering). These methods attempt to incorporate all available nighttime points into cohesive clusters, implicitly assuming that observations reflect a stable spatial structure. In practice, however, noisy or transient points are forced into cluster formation, leading to inflated error estimates and unstable predictions. Similarly, the \textbf{SciKit} method, which selects the most frequently observed nighttime coordinate, remains highly sensitive to sampling bias. Because all points are treated equally, noise that aggregates in non-home locations can dominate the selection process, leading to inaccurate home estimates. Together, these approaches demonstrate a shared vulnerability: they are highly influenced by noise that does not represent true residential behavior.

\textbf{DBSCAN} and \textbf{A2} demonstrate comparatively improved resilience by incorporating built-in filtering mechanisms. DBSCAN requires a minimum density threshold, effectively excluding isolated noise points, while A2 applies dwell-time constraints to identify meaningful stay locations. These mechanisms prevent all points from being treated as equally informative, allowing both methods to better manage noisy observations. However, this robustness comes at the cost of parameter sensitivity. Small changes in neighborhood radius, minimum points, or dwell-time thresholds can significantly alter performance, limiting generalizability. Additionally, A2’s strict filtering can exclude valid residential signals in sparse scenarios, resulting in inconsistent detection.

In contrast, \textbf{GHOST} maintained stable and comparatively low error across all configurations. Rather than relying on clustering or point frequency, GHOST aggregates \textit{stay time} within spatial grid cells, capturing sustained residential behavior while naturally down-weighting transient observations. Its grid-based framework also eliminates the need for clustering entirely, allowing activity to be evaluated within a fixed spatial structure. By avoiding the need to impose structure on noisy data, GHOST is less sensitive to irregular sampling and behavioral variability as seen in the BostonWalks dataset.

The GeoTracker dataset provides a complementary evaluation under controlled conditions, characterized by smaller sample sizes, shorter observation periods, and reduced noise. As expected, all algorithms demonstrated improved performance, with substantial reductions in both MAE and RMSE compared to BostonWalks. For example, the highest RMSE observed in BostonWalks (1,738,379 for A1) dropped to 76,245 in GeoTracker, while the maximum MAE decreased from 452,599 to 27,102 for A2, illustrating the strong influence of data quality on model performance.

Under these cleaner conditions, some methods show notable improvement. For example, \textbf{A1} performs significantly better, as its mean-shift clustering is no longer forced to interpret large amounts of noisy or irregular data as meaningful structure. With fewer spurious points, cluster formation becomes more stable, leading to improved accuracy. In contrast, \textbf{DBSCAN} shows a relative decline in performance. Because it relies on minimum density thresholds to form clusters, the sparser nature of the GeoTracker dataset reduces the number of points available to meet these thresholds, limiting its ability to identify consistent home locations.

Despite these changes, an important finding emerges: strong performance under controlled conditions does not necessarily translate to robustness in real-world mobility datasets. The improved performance observed for DBSCAN, K-MEANS++, A1, A2, and SciKit-mobility was largely driven by favorable data conditions rather than inherent methodological resilience. The GeoTracker dataset spans approximately two weeks per user and contains substantially less noise, irregular sampling, and behavioral variability than BostonWalks. As a result, these methods appear more stable, but their performance may not generalize well to real-world, large-scale mobility data. Even under these favorable conditions, however, GHOST continued to outperform competing methods and maintained consistent performance across both data conditions, demonstrating that its effectiveness is not solely dependent on data quality. Whether applied to relatively clean or highly noisy data, GHOST produced stable and accurate home location estimates, showcasing its robustness and suitability for real-world mobility analysis.

\subsection{GHOST Design Advantages}

GHOST’s performance is driven by several key design components that enhance both accuracy and robustness. First, the use of \textit{stay-time} as the primary behavioral signal enables the algorithm to capture sustained residential occupancy rather than relying on point density or visit frequency. Because GPS sampling rates can vary substantially across users and devices, methods based purely on point counts may incorrectly prioritize noisy or transient locations. Stay-time instead reflects duration of presence, providing a more reliable indicator of residence.

Second, the grid-based framework eliminates the need for clustering operations entirely. Traditional clustering approaches require assumptions regarding cluster structure, density thresholds, and data aggregating around significant locations, all of which become unstable in noisy or irregular mobility datasets \citep{chen2014, wang2023, luo2017, zheng2010}. By dividing the environment into fixed spatial units and aggregating temporal activity within each cell, GHOST avoids these assumptions and limitations.

Third, the hierarchical refinement step improves spatial precision without sacrificing robustness. After identifying the grid cell with the greatest accumulated stay-time, GHOST subdivides the selected cell into smaller bins and identifies the densest local region of activity. This procedure reduces the influence of GPS scatter and local clustering artifacts while preserving stability under sparse observations.

Finally, GHOST's built-in weekend fallback mechanism makes it flexible for incomplete data conditions. Many home detection methods assume that sufficient nighttime observations are always available, which may not hold for users with irregular schedules, night-shift work, inconsistent device usage, or incomplete temporal coverage. GHOST addresses this limitation by shifting to weekend daytime observations when nighttime data are unavailable. 

To directly test the weekend fallback mechanism, all nighttime observations were removed from both testing sets after keeping the optimized parameter configuration fixed (22:00–06:00; grid size = 50 m), forcing GHOST to rely exclusively on weekend daytime data (8 AM–8 PM). Under this setup, grid size becomes the only relevant parameter. On the GeoTracker dataset, two users lacked sufficient weekend coverage after nighttime removal, reflecting the dataset’s sparse temporal structure. For the remaining users, GHOST achieved an MAE of 13.907 m and an RMSE of 17.162 m. Compared to nighttime-based inference (MAE = 22.33 m; RMSE = 35.30 m), errors decreased by 8.42 m (MAE) and 18.14 m (RMSE), indicating that the fallback mechanism not only preserves but can improve performance under controlled conditions. On the BostonWalks dataset, GHOST achieved an MAE of 382.102 m and an RMSE of 1,300.212 m under weekend-only conditions. Relative to nighttime-based results (MAE = 786.53 m; RMSE = 1,938.65 m), this corresponds to reductions of 404.43 m (MAE) and 638.44 m (RMSE). Despite the inherent noise and variability of this large-scale dataset, the fallback mechanism remains effective and reduces extreme errors.

\subsubsection{Parameter Sensitivity}

To further assess robustness, we evaluated sensitivity to parameter selection through a grid size analysis on the GeoTracker dataset using resolutions from 1 m to 250 m (1, 5, 10, 20, 50, 150, 250) and nighttime window configurations with start times between 20:00–22:00 and end times between 05:00–07:00. Performance remained remarkably stable across all configurations, with MAE varying between 22.332 m and 23.048 m and RMSE between 36.574 m and 38.064 m. 

Although slightly lower errors were observed at specific grid sizes (e.g., 50 m and 150 m), these differences were minimal, indicating low sensitivity to parameter selection. Moderate grid sizes, specifically 20 m, 50 m, and 150 m, consistently performed well, suggesting they represent practical default choices. Table \ref{tab:ghost_sensitivity_summary} summarizes these results, where MAE reflects average error, RMSE emphasizes larger deviations, and “best” versus “average” values capture optimal and overall performance across configurations.

\begin{table}[htbp]
\centering
\caption{Sensitivity analysis of GHOST on the GeoTracker dataset. Best-performing configuration and average performance across nighttime windows are shown for each grid size.}
\label{tab:ghost_sensitivity_summary}
\begin{tabular}{lcccc}
\hline
\textbf{Grid Size (m)} & \textbf{Best MAE} & \textbf{Best RMSE} & \textbf{Avg MAE} & \textbf{Avg RMSE} \\
\hline
1   & 22.535 & 37.771 & 22.842 & 37.978 \\
5   & 22.420 & 37.585 & 23.088 & 37.832 \\
10  & 23.035 & 38.056 & 23.048 & 38.064 \\
20  & 22.765 & 36.903 & 22.918 & 37.153 \\
50  & 22.719 & 36.574 & 22.862 & 36.948 \\
150 & 22.332 & 36.980 & 22.352 & 37.005 \\
250 & 22.591 & 37.260 & 22.686 & 37.355 \\
\hline
\end{tabular}
\end{table}

Overall, these results collectively demonstrate that GHOST is robust across both parameter choices and data availability conditions. The model maintains stable performance across a wide range of grid sizes, while the fallback mechanism ensures resilience when primary nighttime signals are missing. Notably, performance is preserved—and in some cases improved—under weekend-only conditions, even in noisy, large-scale datasets. Moderate grid sizes (e.g. 50 m) consistently provide a strong balance between accuracy, stability, and spatial resolution, and are therefore recommended as practical default choices for real-world applications. Together, these findings demonstrate that GHOST provides a reliable and adaptable framework for home detection, supporting its use in mobility data where completeness, consistency, and signal quality cannot be guaranteed.

\section{Conclusion}

This paper presents the development, implementation, and validation of an open-source home detection algorithm, GHOST, designed to infer residential locations from mobile GPS data with high spatial accuracy and strong resilience to data irregularities. Through comprehensive sensitivity analyses and direct comparison with widely used clustering algorithms (All-time clustering, Stay-point method, DBSCAN, K-Means++, and SciKit-mobility home detection), we demonstrate that GHOST consistently achieves lower spatial error, greater robustness to parameter variation, and improved reliability under diverse data conditions. The results highlight the limitations of traditional clustering- and frequency-based approaches when applied to noisy, irregular, and behaviorally complex mobility datasets, where sensitivity to parameter selection, GPS noise, and sparse observations can substantially degrade performance. By leveraging time-based aggregation, a flexible grid-based framework, adaptive temporal filtering, and a hierarchical refinement step, GHOST effectively captures behaviorally meaningful signals of residential activity while remaining stable in the presence of noise and incomplete observations. The built-in weekend fallback mechanism further improves robustness under irregular schedules and incomplete nighttime coverage, allowing the algorithm to maintain strong performance even when primary temporal signals are unavailable. Its ability to operate effectively with limited data, accommodate diverse behavioral patterns, and deliver accurate results with minimal parameter tuning makes it well suited for a wide range of mobility research and practical decision-making applications. Future work will focus on further improving GHOST’s performance through the integration of external validation layers, particularly by incorporating parcel-level land-use data to verify that inferred home locations correspond to residential land uses rather than commercial or institutional areas. This additional verification step offers a promising direction for reducing misclassification and enhancing the reliability and interpretability of home detection in large-scale mobility datasets.

\section{Acknowledgments}
This material is based upon work supported by the U.S. National Science Foundation (NSF) under awards Nos. 2338959 and 2416202, as well as the BostonWalks Dataset provided by the Massachusetts Institute of Technology (MIT). Any opinions, findings, conclusions, or recommendations expressed in this material are those of the authors and do not necessarily reflect the views of the NSF or MIT. During the preparation of this work, the authors used ChatGPT to assist with grammar review and language refinement. After using this tool, the authors carefully reviewed and edited the content as needed and take full responsibility for the final publication.

\newpage

\appendix
\section{Appendix}
\label{app:geotracker}

\begin{table}[htbp]
    \centering
    \caption{Summary of GPS Data Collection for GeoTracker Users}
    \label{tab:user_gps_summary}
    \resizebox{\textwidth}{!}{%
    \begin{tabular}{|c|c|c|c|c|c|c|c|c|}
        \hline
        \textbf{User} & \textbf{Total GPS Points} & \textbf{Days Collected} & \textbf{Avg Points/Day} & \textbf{Min Points} & \textbf{Max Points} & \textbf{Start Date} & \textbf{End Date} & \textbf{Coverage (\%)} \\ \hline
        1 & 13420 & 11 & 1220 & 238 & 3136 & 06/19/2025 & 06/29/2025 & 57.20 \\ \hline
        2 & 66671 & 29 & 2299 & 2 & 19924 & 06/15/2025 & 07/14/2025 & 41.52 \\ \hline
        3 & 12984 & 9 & 1443 & 1 & 7134 & 06/17/2025 & 06/27/2025 & 53.24 \\ \hline
        4 & 32828 & 15 & 2189 & 486 & 11348 & 08/19/2025 & 09/02/2025 & 57.78 \\ \hline
        5 & 16303 & 5 & 3261 & 43 & 11059 & 06/17/2025 & 06/26/2025 & 51.67 \\ \hline
        6 & 8765 & 11 & 797 & 6 & 3242 & 06/27/2025 & 07/08/2025 & 54.17 \\ \hline
        7 & 4865 & 14 & 348 & 1 & 1132 & 06/23/2025 & 09/11/2025 & 30.06 \\ \hline
        8 & 21622 & 9 & 2402 & 30 & 4959 & 07/02/2025 & 07/17/2025 & 50.46 \\ \hline
        9 & 10537 & 4 & 2634 & 43 & 6045 & 06/25/2025 & 07/05/2025 & 39.58 \\ \hline
        10 & 8047 & 8 & 1005 & 26 & 2205 & 02/03/2026 & 02/13/2026 & 31.33 \\ \hline
    \end{tabular}%
    }
\end{table}

\bibliographystyle{apalike}

\newpage
\bibliography{References}
\end{document}